# Overlay-aware Variation Study of Flip FET and Benchmark with CFET

Wanyue Peng, Haoran Lu, Jingru Jiang, Jiacheng Sun, Ming Li, Runsheng Wang, Heng Wu†, Ru Huang
School of Integrated Circuits, Peking University, Beijing 100871, China, †email: hengwu@pku.edu.cn

## Abstract

In this work, we carried out an overlay-aware variation study on Flip FET (FFET) considering the impact on RC parasitics induced by lithography misalignment in the backside processes, and benchmarked it with CFET in terms of the power-performance (PP) and variation sources. The iso-leakage frequency degrades up to 2.20% with layout misalignment of 4 nm. It's found that the Drain Merge resistance degrades significantly with misalignment increasing and is identified as the major variation sources. Through careful DTCO with design rules optimization, the variation can be greatly suppressed, while the resistance fluctuation of the DM also drops substantially. Monte Carlo random experiments were also conducted, validating the variation reduction. Comparing with the CFET featuring self-aligned gate and much less overlay induced misalignment, fortunately, FFET's PP is still better except when misalignment reaches 8 nm, which is out of spec and nearly impossible. Considering the variabilities induced by the high aspect ratio processes, CFET still faces big challenges compared with FFET.

## Introduction

As the traditional scaling coming to an end, the Flip FET (FFET) was promoted as a promising solution for future logic technology [1]. However, normal FFET only features self-aligned active (Fig. 1(a)), leaving the backside (BS) layers possibly misaligned to frontside (FS) ones, which is different from CFET [2][3] with naturally self-aligned gate, as shown in Fig. 1(b).

Design of experiments (DoEs) based on misalignment assumption has been conducted to investigate how the misalignment affects, followed by careful variation source validation and DTCO. Further power-performance (PP) analysis between the FFET and the CFET proves that FFET could still be a better solution than the CFET despite its misalignment-induced PP degradation, while CFET has more variation sources due to high aspect ratio processes. Note that the overlay impact on the intrinsic device performance is not yet considered in this work, deserving further investigation.

## Variation Sources Exploration

The FFET [1] is composed of back-to-back-stacked transistors with only the active formed in a self-aligned manner. As the first lithography on the backside is the BS-Gate, it's assumed that misalignment mainly occurred between BS-Gate and FS-Gate, while other BS masks would align to BS-Gate. In this case, the intrinsic performance of FS and BS devices keeps unchanged in this work.

To describe the misalignment distance and directions more precisely, we introduce the Misalignment Vector along which the BS masks shift relative to the FS layers, as illustrated on Fig. 2. Based on backside EUV lithography experiment [4], 0 ~ 4 nm is assumed to be the most reasonable misalignment range between FS and BS layers. Under each misalignment condition, a 15-stage ring-oscillator with fan-out 3 was used for PP analyses at iso-leakage @VDD = 0.7 V, and the results are displayed on Fig. 3. Only frequency variation is shown as the power variation is too small compared with the frequency. The maximum frequency degradation is up to 2.20% when the misaligned BS layers move along the Y-axis to a very negative place.

To lower the variation, the main source of frequency fluctuation should be identified first. Specially, the Drain Merge (DM) and the Gate Merge (GM) are formed on the BS S/D region and gate region respectively to connect BS and FS transistors. Once there is a misalignment, the DM and GM cannot directly land on frontside transistor's metalized drain (MD) and gate metal precisely (Fig. 4). Therefore, the DM and the GM have the largest structural variations, which influence their parasitic resistances a lot. It's assumed that the frequency variation mainly derives from the resistance variation of the DM and the GM and developed two models. One is the DM-only model, which only considers the resistance variation of the DM and replaces the DM resistance sub-netlist in the baseline Inverter (INV) netlist with DM resistance sub-netlist under various misalignment conditions. The other one is the GM-only model which follows a similar approach. After comparing different converted frequency with experimental frequency, it's validated that the resistance of the DM contributes to the majority of the frequency variation (Fig. 5).

The reason why the variation in GM resistance ($\Delta R_{GM}$) has a smaller influence compared to DM is that the $\Delta R_{GM}$ is very small relative to the total resistance of the gate (less than 10%), so the impact of $\Delta R_{GM}$ is minimal. Additionally, it should be acknowledged that the parasitic variation under misalignment varies with different design rules. In the baseline design rule, the DM is mainly buried in the Gate Cut and the overlay between the DM and the gate is small (as shown in Fig. 6(a)), so the DM contributes little to $C_{gd}$. As a result, the $\Delta C_{gd}$ matters less than $\Delta R_{DM}$.

## DTCO and Monte Carlo Validation

Previous work has confirmed that the DM determines the frequency variation. So to reduce the variation, lowering the variation of the DM resistance ($\Delta R_{DM}$) is needed. Further investigation into the baseline design rule reveals that the DM would be blocked by the S/D epi (Fig. 6(a), taking INV as an example), thus the size of DM varies significantly. To leave more margin between the Drain Merge and the active, the power rail can be moved to other side of the layout thanks to the flexibility of the FFET layout design, making the Drain Merge farther away from the active, as shown in Fig. 6(b). The MD was also enlarged at the same time while all the other structures kept the same. The $\Delta R_{DM}$ under the baseline design rule and the optimized design rule were also extracted. The standard deviation of $\Delta R_{DM}$ decreases 97.2% compared with the baseline one, which matches the expectation. Misalignment-induced frequency variation is suppressed evidently after the design rule optimization (Fig. 8) compared to the baseline in Fig. 3. The maximum frequency degradation is optimized from 2.20% to 1.30% due to the $\Delta R_{DM}$ improvement.

In the experiments above, the probabilities of different Misalignment Vectors occurring across the wafer is not considered, which actually is not the same. To mimic the real case, It's assumed that the probability of the Misalignment Vector decreases as the misalignment distance increases and most of the Vectors exist within the range of 0-4 nm. The misalignment distribution assumption is depicted in Fig. 9. Based on this distribution, 10,000 Monte Carlo simulations were conducted, randomly selecting a Vector each time and calculating the corresponding frequency. Then the misalignment-induced frequency distributions under the baseline design rule and the optimized design rule were extracted, as shown in Fig. 10(a) & (b), respectively. The standard deviation of frequency after DTCO decreases by 19.7% compared with the baseline one.

All the elements above should be considered into new design rule enablement, and the variation of $C_{gd}$ should also be considered.

## Benchmark with CFET

In the previous work [1], it's proved that the 3.5T dual-fin FFET outperforms the 4T dual-fin CFET by using RO simulation. For here, the power-performance degradation derived from the overlay in the FFET is evaluated, with worst cases under each misalignment range selected. The results in Fig.11 show that the PP of the FFET is weaker than that of the CFET only when the misalignment distance reaches 8 nm, while the situation is considered impossible if the EUV lithography is used[4]. Both FFET and CFET are under tight design rules (with short distance between vias to active, small gate extension [4], etc.). So even though there is some misalignment-induced degradation in the FFET, its performance can still outperform the CFET.

Misalignment between frontside FEOL layers and backside FEOL layers is a unique source of variation in FFET. But CFET is free of this issue because its structures are mostly formed on the frontside of the wafer in a self-aligned manner. However, the unique high aspect ratio processes in CFET could also introduce significant variations [4-8], as shown in Fig. 12.

Table. 1 gives an analysis of these variations. The wafer flip process is considered the reason for patterning misalignment. But it also greatly reduced the numbers of high aspect ratios (HAR). The CFET can also replace some HAR processes at the cost of introducing flip process, like the bottom MD formation and BPR formation can be substituted by direct backside contact [9-13]. As the influence of patterning-induced fluctuation of FFET is limited while the high aspect ratio processes matter significantly, it's indicated that the FFET can still exceed the CFET even in variation aspect.

## Conclusion

In this work, we conducted an overlay-aware variation study on FFET using advanced DTCO flow, in a parasitic perspective. A Precise model was established to verify the DM resistance as the main variation source. Based on it, new design rule was proposed and validated with much lower the frequency variation. Furthermore, to comprehensively study the frequency distribution, Mento Carlo random experiments were also conducted. Power-performance comparison and the trade-off between flip-induced patterning misalignment and high aspect ratio processes variations were analyzed, implying that FFET retains great advantages over CFET.

## Acknowledgments

All the authors gratefully acknowledge the funding support from National Key Research and Development Program of China under Grant 2023YFB4402200; and in part by the 111 Project under Grant 8201702520.

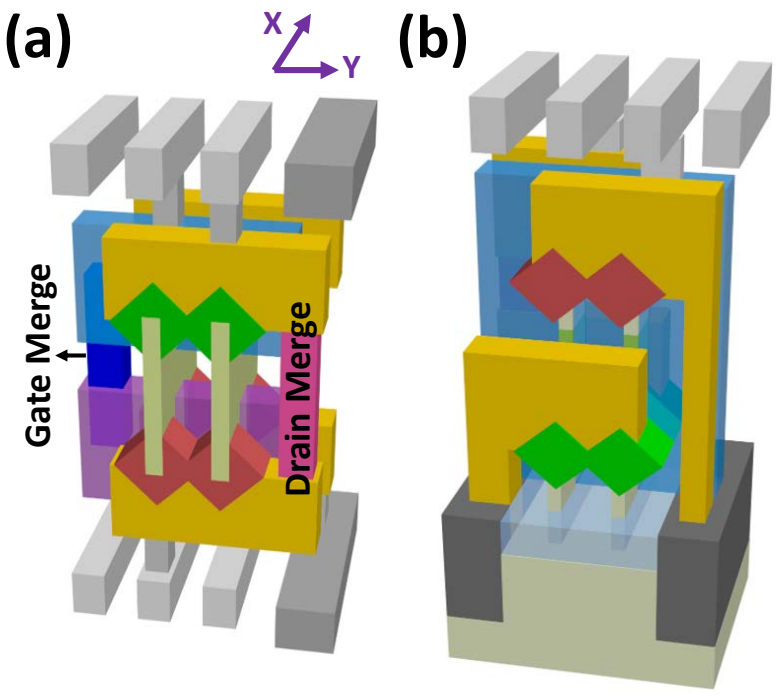

**Fig. 1** (a) 3D schematic of the 3.5T dual-fin FFET. (b) 3D schematic of the 4T dual-fin CFET with BPR.

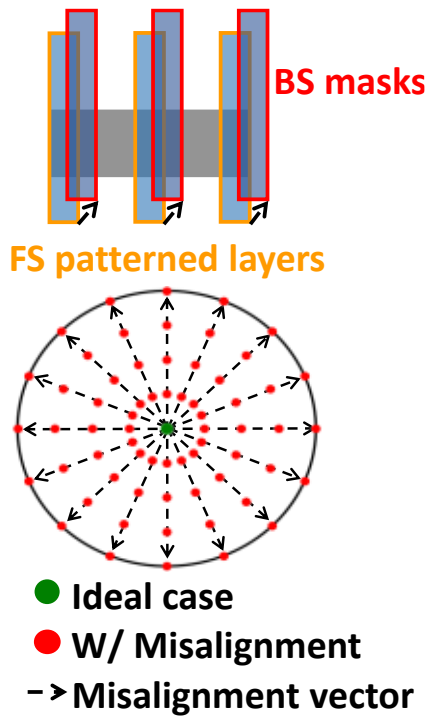

**Fig. 2** The Misalignment of the FFET. BS masks shift along the Misalignment Vectors relative to the FS layers.

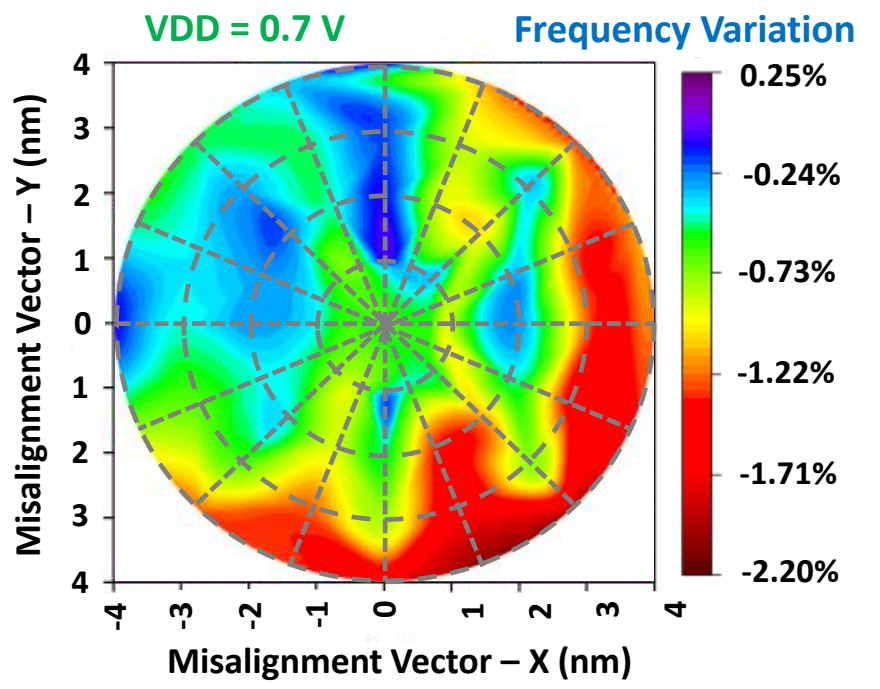

**Fig. 3** Misalignment-induced FFET frequency variation distributed along all directions with the maximum possible displacement assumed to be 4 nm.

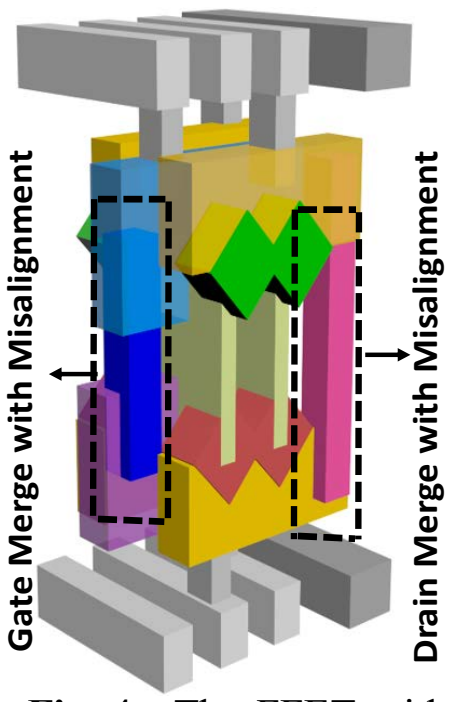

**Fig. 4** The FFET with misalignment where the DM and the GM have obvious mismatches.

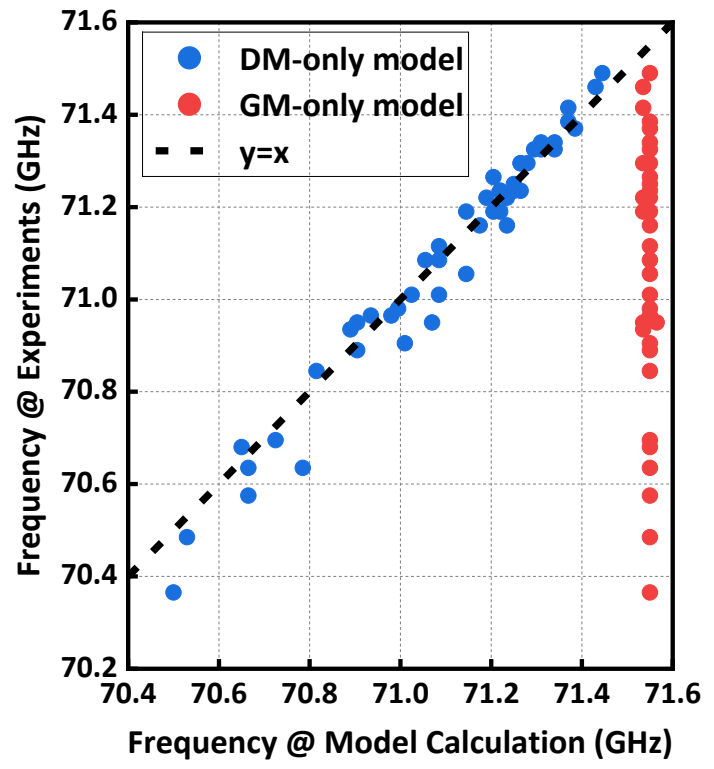

**Fig. 5** Frequency under misalignment experiments vs. frequency converted by replacing netlists without misalignment with DM or GM resistance sub-netlists.

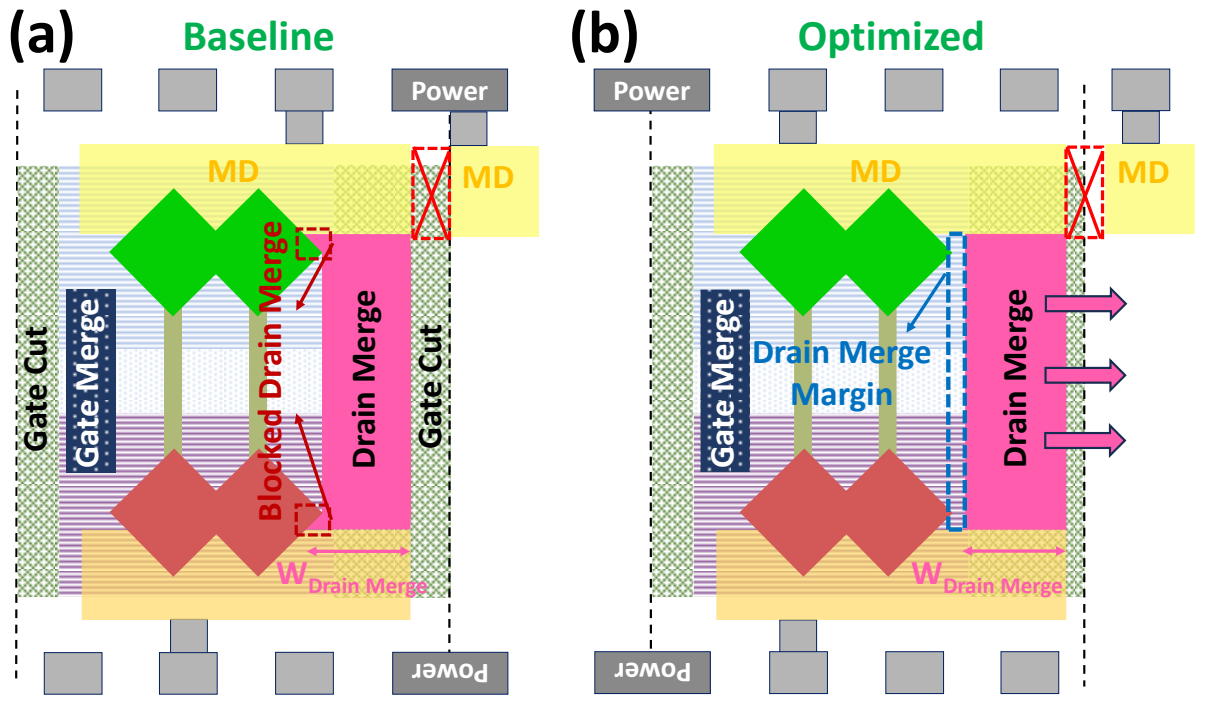

**Fig. 6** (a) Baseline design rule of FFET. The DM is blocked by the epi with misalignment. (b) The optimized design rule with better margins between the DM and the active. The MD can be closer to the boundary after modifying the power rail, thus the same for the DM.

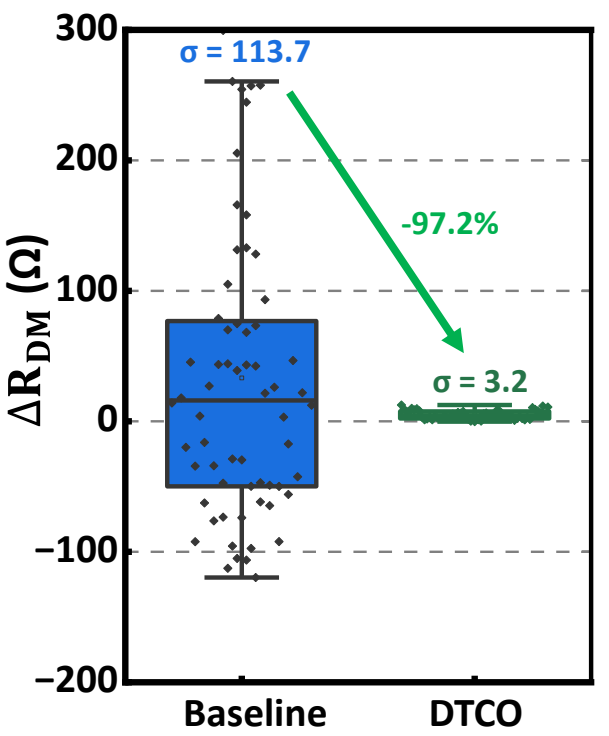

**Fig. 7** $\Delta R_{DM}$ under the baseline and optimized design rule. The standard deviation of $\Delta R_{DM}$ decreases by 97.2%.

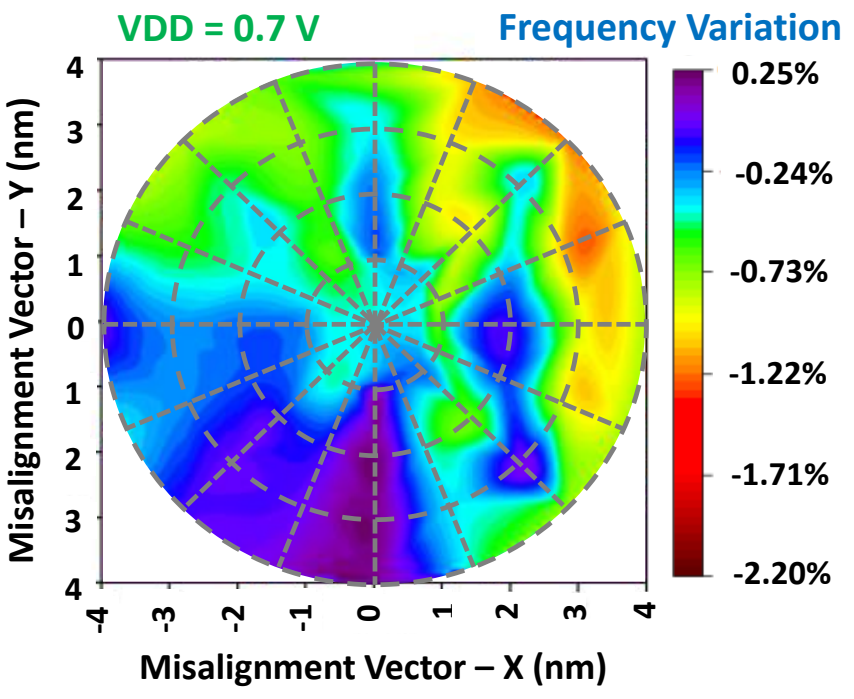

**Fig. 8** Misalignment-induced FFET frequency variation distributed along all directions after DTCO. Compared to baseline case in Fig. 3, the frequency variation is suppressed evidently.

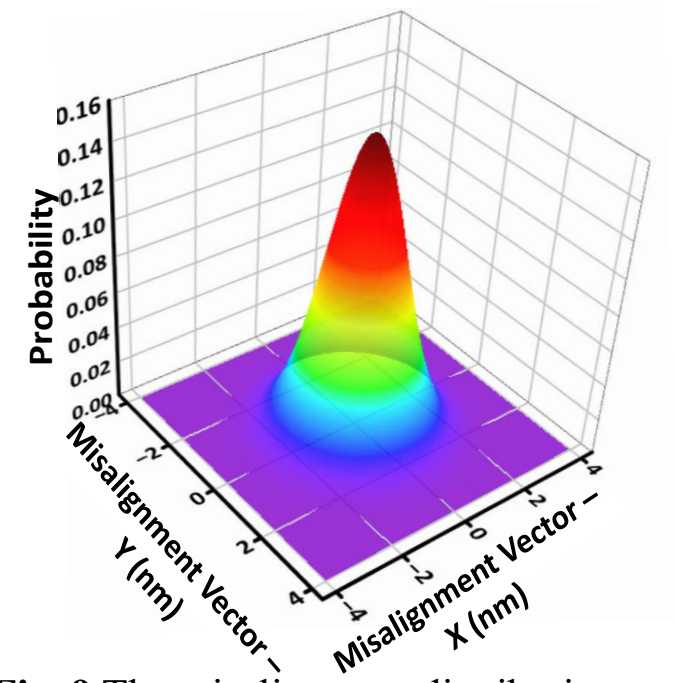

**Fig. 9** The misalignment distribution is assumed to follow a two-dimensional normal distribution. The cumulative probability within the range of 0-4 nm is 99.7%.

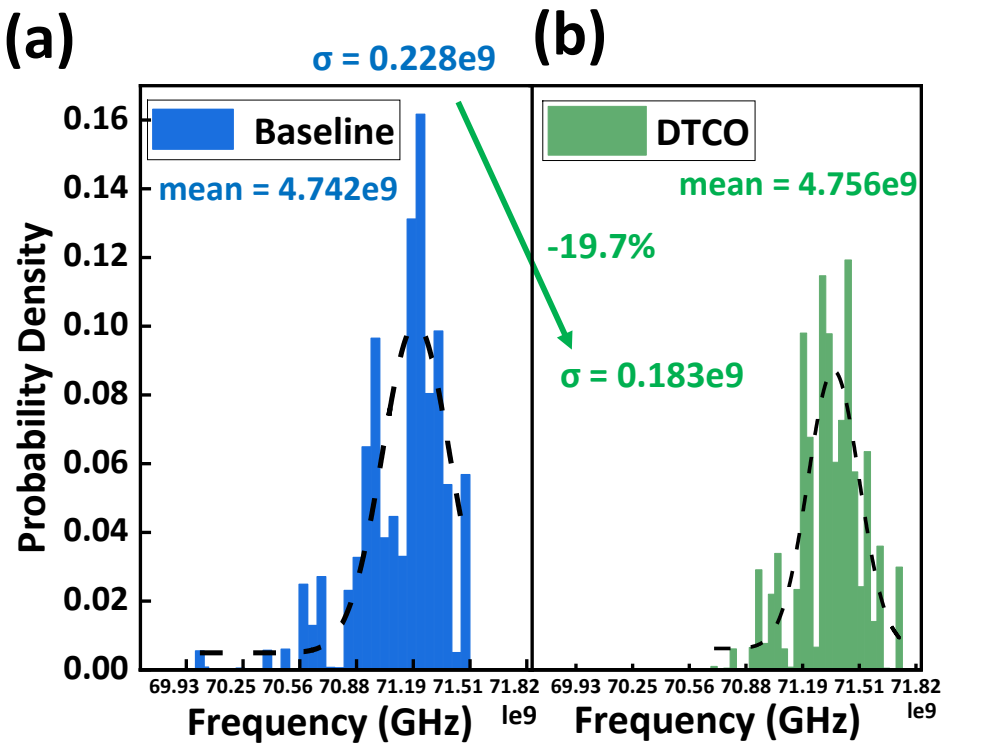

**Fig. 10** Monte Carlo experiments on frequency occurrence performed 10,000 times, for baseline (a) and DTCO (b), based on misalignment vectors distribution in Fig. 6. The standard deviation decreases by 19.7%.

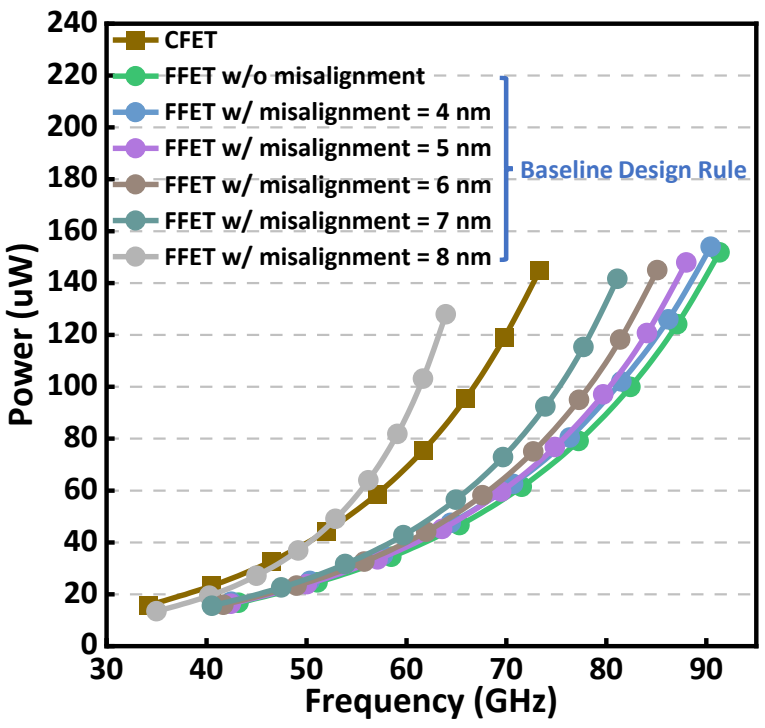

**Fig. 11** Power-Performance curves of the CFET's and the FFET's baselines and worst cases of different misalignment conditions.

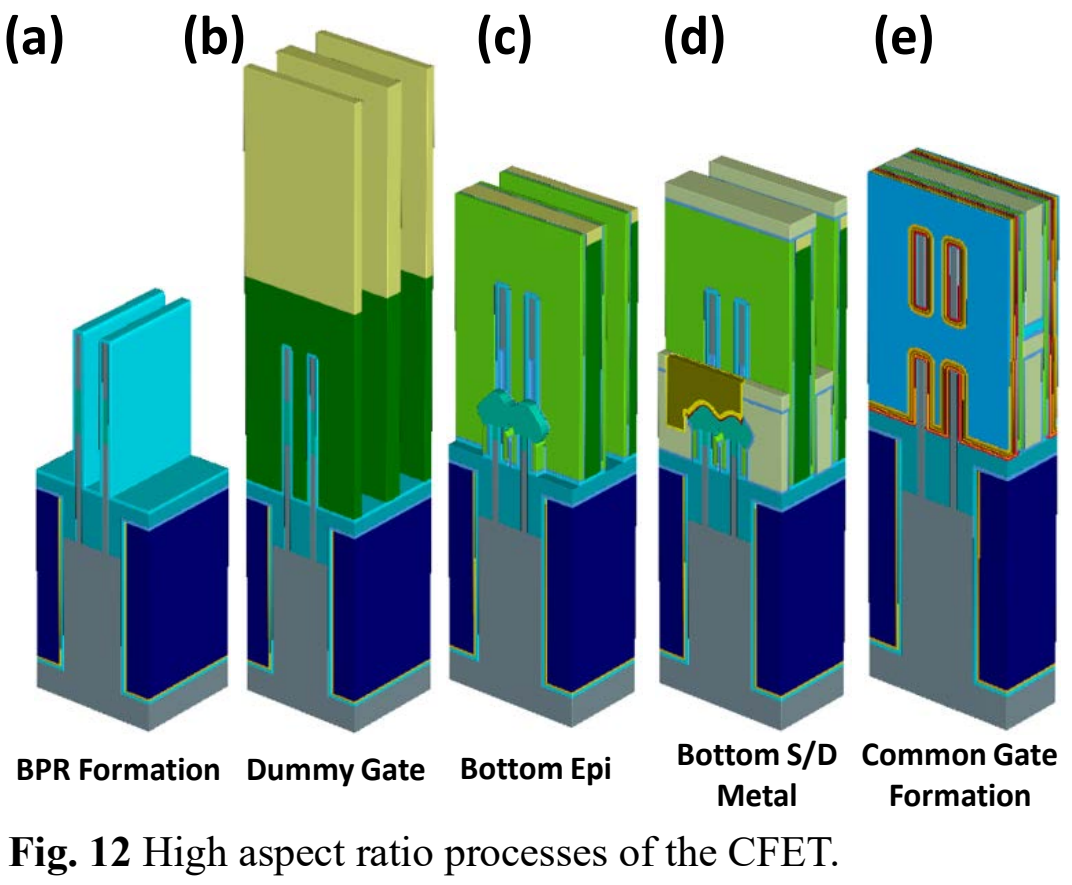

**Fig. 12** High aspect ratio processes of the CFET.

| Variation Sources | F2ET | CFET | Description |
| --- | --- | --- | --- |
| Lithography | Yes | No | Wafer flip induced misalignment in the FFET. |
| BPR formation | No | Yes | A deep trench should be etched on the wafer frontside in the CFET. |
| Dummy gate | No | Yes | Dummy gate is formed on wafer frontside for both top and bottom FETs in the CFET. |
| Epi formation | No | Yes | Bottom epi & top epi & metal gate & MD are all formed on the frontside of the wafer in the CFET, while they are not in the FFET. |
| RMG | No | Yes | |
| MD formation | No | Yes | |

**Table 1** Variation sources from lithography and HAR processes in FFET and CFET.